\documentclass[
  reprint, 
  superscriptaddress,
  amsmath,amssymb,
  aps,
  prl,
  longbibliography
]{revtex4-2}

\usepackage[utf8]{inputenc}
\usepackage[T1]{fontenc}

\usepackage{dsfont}
\usepackage{mathtools} 
\usepackage{bm,bbm,MnSymbol}
\usepackage{physics} 

\usepackage{graphicx}
\usepackage{dcolumn}
\usepackage{booktabs}
\usepackage[svgnames]{xcolor} 

\usepackage[top=2cm, bottom=2cm, left=1.5cm, right=1.5cm]{geometry}
\usepackage{lipsum}
\usepackage[mathlines]{lineno}

\usepackage{orcidlink}

\usepackage{hyperref}
\hypersetup{
    colorlinks=true,
    citecolor=blue,
    linkcolor=blue,
    filecolor=blue,  
    urlcolor=DarkBlue,
    pdftitle={Bell Tests notes},
    pdfpagemode=FullScreen,
}
\begin{document}
\preprint{APS/123-QED}
\title{Bell Test of Photons from Electron-Positron Annihilation \\ via POVM-based Compton Polarimetry}

\author{Jack Clarke\,\orcidlink{0000-0001-8055-449X}}
 \email{jack-clarke@ucl.ac.uk}
 \affiliation{Department of Physics and Astronomy, \href{https://ror.org/02jx3x895}{University College London}, Gower Street, WC1E 6BT London, United Kingdom}
\author{Preslav Asenov\,\orcidlink{0009-0003-4232-4521}}
 \email{preslav.asenov.20@ucl.ac.uk}
 \affiliation{Department of Physics and Astronomy, \href{https://ror.org/02jx3x895}{University College London}, Gower Street, WC1E 6BT London, United Kingdom}
\author{Jesse Smeets\,\orcidlink{0009-0009-5205-4936}}
\email{j.smeets2@vu.nl}
\affiliation{Department of Physics and Astronomy, LaserLaB, \href{https://ror.org/008xxew50}{Vrije Universiteit Amsterdam}, De Boelelaan 1100, Amsterdam, 1081 HZ, The Netherlands}
\affiliation{Department of Applied Physics, \href{https://ror.org/02c2kyt77}{Eindhoven University of Technology}, P.O. Box 513, 5600 MB Eindhoven, The Netherlands}
\author{Jia-Shian Wang\,\orcidlink{0000-0003-2499-7039}}
\email{josh.wang@ucl.ac.uk}
\affiliation{Department of Physics and Astronomy, \href{https://ror.org/02jx3x895}{University College London}, Gower Street, WC1E 6BT London, United Kingdom}
\author{David B. Cassidy\,\orcidlink{0000-0001-8332-5553}}
 \email{d.cassidy@ucl.ac.uk}
 \affiliation{Department of Physics and Astronomy, \href{https://ror.org/02jx3x895}{University College London}, Gower Street, WC1E 6BT London, United Kingdom}
\author{Alessio Serafini\,\orcidlink{0000-0002-4509-7470}}
 \email{serale@theory.phys.ucl.ac.uk}
 \affiliation{Department of Physics and Astronomy, \href{https://ror.org/02jx3x895}{University College London}, Gower Street, WC1E 6BT London, United Kingdom}
\date{\today}
\begin{abstract}

Quantum entanglement between gamma-ray photons emitted following electron-positron annihilation is expected to be maximal and may be characterized via non-classical polarization correlations. However, this is difficult to verify experimentally because there are no established schemes that approach ideal projective-polarization measurements for high-energy photons. Hence, polarization entanglement between MeV-scale annihilation photons has not yet been conclusively demonstrated. We develop here a framework that models polarization measurements of high-energy photons via Compton polarimetry, employing the formalism of positive operator-valued measures (POVMs). We extend the POVM description to sequences of repeated interactions and show that the measurement converges toward an ideal projective measurement of linear polarization as the number of interactions increases. We demonstrate that this progressive improvement in measurement sharpness can enable the experimental violation of CHSH inequalities.
\end{abstract}

\maketitle
\textit{Introduction.}---Since Wheeler's 1946 proposal~\cite{wheeler1946polyelectrons}, witnessing polarization entanglement of photons from electron-positron annihilation has remained an elusive goal of quantum electrodynamics~\cite{wu1950theangular,bohm1957discussion,clauser1978bell}. 
Bell repeatedly advocated for the study of polarization correlations of these annihilation photons to test for violations of his eponymous inequality~\cite{bell1964onthe,bell2004speakable}. 
However, standard polarizers are effectively transparent to $0.5\,\text{MeV}$ photons. This forces a reliance on Compton polarimetry which, in its standard form, amounts to an imperfect non-projective measurement of photon polarization that yields correlations too weak to violate a Bell inequality~\cite{clauser1969proposed}.
These insufficient correlations are precisely why the early effort of Kasday, Ullman, and Wu (KUW)~\cite{kasday1971experimental,kasday1972distribution,kasday1975angular}, along with more recent iterations~\cite{wilson1976measurement,bruno1977measurement,osuch1996experimental,ivashkin2023testing}, could not produce, \textit{even in principle}, a definitive Bell violation.
Currently, there are no experiments that have certified polarization entanglement at the MeV scale, in stark contrast to the optical domain, where polarizers are straightforward to produce and Bell tests have proved plentiful~\cite{freedman1972experimental,aspect1982experimentalrealization,aspect1982experimental,weihs1998violation,brunner2014bell}.

Despite this persistent experimental challenge, harnessing the polarization entanglement of annihilation photons promises to be a powerful resource for both applied technologies and fundamental research.
Exploiting polarization entanglement would allow one to reduce background errors in the next generation of PET scanners~\cite{toghyani2016polarisation,watts2021photon,moskal2025nonmaximal}, and studying the quantum information properties of annihilation photons from positronium decay would impact precision tests of QED~\cite{skalsey1991first,karshenboim2004precision,yamazaki2010search,cassidy2018experimental} and the search for new physics beyond the standard model~\cite{adkins2022precision}.   
To unlock these applications, a rigorous and efficient treatment of QED scattering processes is required for calculating information-theoretic quantities. For this purpose, one can resort to the well-established Stokes--Mueller matrix method, which allows tree-level Compton scattering to be described as a linear transformation of Stokes parameters~\cite{wightman1948note,fano1949remarks}. Crucially, this framework has been generalized to describe sequential Compton interactions~\cite{mcmaster1961matrix,caradonna2024stokes,caradonna2024kinematic}, recently enabling the experimental description of a triple Compton scattering process to study decoherence~\cite{bordes2024first}. 
However, while previous experiments have provided evidence compatible with entanglement~\cite{watts2021photon,bordes2024first,moskal2025nonmaximal}, they do not conclusively prove it.
As we detail below, existing works do not adopt the quantum-information-theoretic definition of an entanglement witness~\cite{supp}. Furthermore, they ultimately cannot certify entanglement as they are constructed from the Pryce--Ward cross section~\cite{pryce1947angular}, which can be reproduced by a local hidden variable (LHV) model~\cite{kasday1971experimental,kasday1972distribution,pei2026can}. 

Here, we use the tools of quantum information theory to enhance standard Compton polarimetry by including sequential scattering events, predicting a definitive violation of the CHSH inequality~\cite{clauser1969proposed}. From the Klein--Nishina differential cross section for an arbitrary state of photon polarization~\cite{klein1928scattering}, we derive the positive operator-valued measure (POVM) corresponding to a single Compton scattering event followed by photon detection. 
We then extend this description using the Stokes--Mueller formalism to describe sequential scattering events, which allows the original KUW experiment to be generalized to arbitrary scattering sequences.
We show that by utilizing two or more scattering events, thus going beyond the Pryce--Ward cross section, the LHV bound of the CHSH inequality can be violated by optimizing over scattering angles. 
Furthermore, we show that as the number of scattering events increases, the corresponding POVM converges to an ideal projective measurement, and the Tsirelson bound is approached~\cite{cirel1980quantum}.

\begin{figure*}
    \centering
    \includegraphics[width=0.9\linewidth]{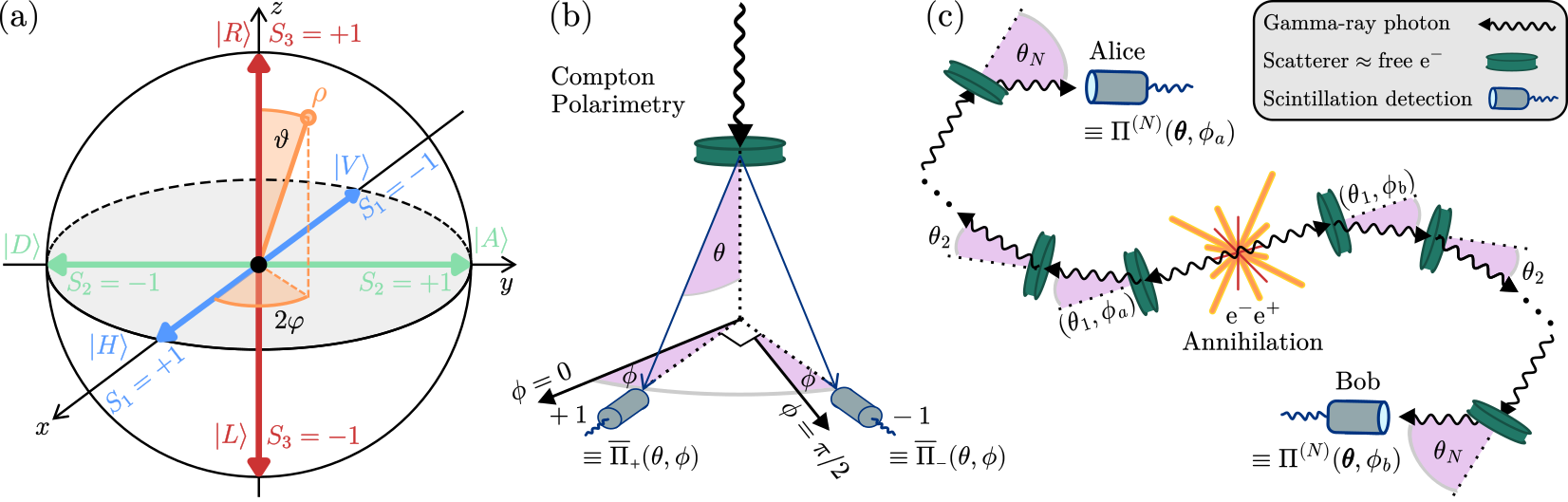}
    \caption{(a) The Bloch sphere in polarization space $(\vartheta,\varphi)$. States along the equator of the Bloch sphere represent linearly polarized states. The north and south poles represent the circularly polarized eigenstates, which Compton polarimetry cannot distinguish. (b) POVM-based Compton polarimetry. Following a Compton scattering event, two polarization-insensitive detectors separated by azimuthal angle $\Delta \phi=\pi/2$ realize non-projective measurements of orthogonal linear polarization states. Detection at arbitrary $(\theta,\phi)$ corresponds to the POVM element $\Pi^{(1)}(\theta,\phi)$. In the KUW Bell test, a value $\pm1$ is assigned to detection at $(\theta,\phi_{\pm})$ with $\phi_{+}=\phi$ and $\phi_{-}=\phi+\pi/2$, corresponding to the dichotomic, normalized POVM elements $\overline{\Pi}_{\pm}(\theta,\phi)$. (c) Setup for the generalized-KUW Bell test experiment of annihilation photons. After $N\geq1$ scattering events per photon---where $N=1$ is the original KUW experiment---detection is described by $\Pi^{(N)}(\bm{\theta},\phi)$ for the co-planar trajectories described below. The dichotomic, normalized POVM elements are $\overline{\Pi}_{\pm}(\bm{\theta},\phi)$ and Alice and Bob calculate the CHSH function from their two measurement settings parametrized by $\phi$: $\{\phi_a,\phi_{a'}\}$ and $\{\phi_b,\phi_{b'}\}$. The $+1$ outcome corresponds to the plane shown here, while $-1$ corresponds to detection out of the plane.}
    \label{fig:setup}
\end{figure*}

These results advance the growing research effort to integrate quantum information theory and high-energy physics. Our POVM-based analysis has a wide range of applications at this new intersection, from witnessing entanglement in different scattering processes~\cite{acin2001three,cervera2017maximal,fedida2023tree,blasone24,blasone25,liu25,smeets2025qedtool} and optimizing quantum estimation~\cite{ignoti2025large,asenov2026quantum,piotrak2026quantum,frugiuele2026quantum}, to investigating the information-theoretic properties of the vacuum~\cite{dudal2023maximal,fedida2024entanglement} and interpreting data from particle collider experiments~\cite{atlas2024observation,cms2024observation,afik2025quantum}. 
%

%
\textit{POVM Analysis of Compton Scattering.}---Here, we discuss Compton polarimetry by deriving its POVM from the Klein--Nishina formula. 
More specifically, given two sets of scattering angles $\{\theta,\phi_{\pm}\}$, labelled with $\pm$, we determine the matrices $\overline{\Pi}_{\pm}$ such that the outcome probabilities are $p_{\pm}(\theta,\phi_{\pm})={\rm Tr}[\rho \overline{\Pi}_{\pm}]$, where $\rho$ is the $2\times2$ normalized density matrix in polarization space of the initial photon~\footnote{Recall a POVM is a set of positive operators $\{\Pi_{\mu}\}$, such that $\sumint_{\mu} \Pi_{\mu}={\mathds{1}}$, so that each measurement outcome occurs with probability (or probability density) $p(\mu)={\rm Tr}[\rho \Pi_{\mu}]$. Such a generalized version of the Born rule is necessary to describe non-projective measurements in quantum mechanics.}.
Thus, we are able to identify the parameter $\beta$ that quantifies the sharpness of the polarization measurement, and grant new significance to it. We then generalize these results to an arbitrary number of sequential Compton scattering events.

Using the fact that the Pauli matrices and the identity form a basis of Hermitian operators, it will be convenient to vectorize the density matrix $\rho$ as a Stokes vector: $\ket{S} = (S_0,S_1,S_2,S_3)^{\sf T}$, with $S_j= {\rm Tr}[\rho \sigma_j]$ for $\bm{\sigma}=\{\mathds{1},\sigma_x,\sigma_y,\sigma_z\}$.
Note that $\rho=\frac{1}{2}\sum_{j}S_j\sigma_j$ is the qubit Bloch vector of the photon prior to scattering, which is illustrated in Fig.~\ref{fig:setup}(a). Here, $\{\ket{H},\ket{V}\}$, $\{\ket{A},\ket{D}\}$, and $\left\{\ket{R},\ket{L}\right\}$ are the eigenvectors of Pauli matrices $\sigma_x$, $\sigma_y$, and $\sigma_z$, respectively.
Importantly, in the following we shall not assume the state remains normalized, i.e. $S_0=1$, after scattering, which allows us to keep track of the probabilities of sequential interactions through the Stokes--Mueller formalism. Hence, after scattering the Stokes vector represents an unnormalized conditional state in polarization space.

The Klein--Nishina formula gives the differential scattering cross section as a function of the polar angle $\theta$ and azimuthal angle $\phi$, defined from the momentum vector of the incoming photon---see Fig.~\ref{fig:setup}(b). Upon normalization, the polarization-dependent Klein--Nishina formula yields the scattering probability density ${\rm d}p(\theta,\phi)$ for any initial normalized state with Stokes vector $\ket{S}$ (with $S_0=1$; the state is only unnormalized after scattering), as
\begin{align}\label{eq:KN_1a}
    {\rm d}p(\theta,\phi) = \mathcal{N}\left[1+ \beta\left( S_1 \cos 2\phi + S_2 \sin 2\phi \right)\right]{\rm d}\Omega,
\end{align}
where $\mathcal{N}=r_e^2 \alpha/\sigma_{\text{tot}}^{(1)}$, $\alpha$ and $\beta$ are functions of $E_0$ and $\theta$ given in the End Matter, $E_0$ is the initial photon energy in units of electron mass ($E_0=1$ corresponds to $511\,\text{keV}$), $r_e$ is the classical electron radius, and $\sigma_{\text{tot}}^{(1)}$ is the total cross section for one Compton scattering.
Comparison of Eq.~(\ref{eq:KN_1a}) and 
\begin{align}\label{eq:dprob}
\frac{{\rm d}p (\theta,\phi)}{{\rm d}\Omega}={\rm Tr}[\rho \Pi^{(1)}(\theta,\phi)]=\frac{1}{2}\sum_{j}S_j {\rm Tr}[\sigma_j \Pi^{(1)}(\theta,\phi)],
\end{align}
which must hold for all $S_1$, $S_2$, and $S_3$, yields the POVM elements in the $\{\ket{L},\ket{R}\}$ basis:
\begin{align}\label{eq:POVM1}
    \Pi^{(1)}(\theta,\phi)=r_e^2\,\dfrac{\alpha(\theta)}{\sigma_{\text{tot}}^{(1)}}\begin{pmatrix}
        1 & \beta(\theta)e^{-2i\phi} \\
        \beta(\theta)e^{2i\phi} & 1
    \end{pmatrix}.
\end{align}
In general, the POVM element above is only normalized if integrated upon scattering angles. 
However, to achieve a substantial simplification, we may consider only two angles $\phi_{+}=\phi$ and $\phi_{-}=\phi+\pi/2$, such that $\Pi_{+}+\Pi_{-} \propto {\mathds1}$. Under this condition, the two elements of the effective, filtered POVM in polarization space are given by the normalized version of Eq.~(\ref{eq:POVM1}) (see the Supplemental Material \cite{supp} for a detailed justification of this step):
\begin{align}\label{eq:POVM2}
    \overline{\Pi}_{\pm}(\theta,\phi)=\frac{1}{2}\begin{pmatrix}
        1 & \pm\beta(\theta)e^{-2i\phi}  \\
        \pm \beta(\theta)e^{2i\phi}  & 1
    \end{pmatrix}.
\end{align}

We have thus distilled Compton polarimetry to the analysis of the $2\times2$ matrices $\overline{\Pi}_{\pm}$. The function $\beta(\theta)$, known as the `Compton analyzing power', does indeed completely parametrize the POVM, and determines how far it is from a perfect projective measurement. The latter are one-dimensional projectors, and hence have determinant zero. Eq.~(\ref{eq:POVM2}) immediately reveals that this is attained for $\beta(\theta)=1$. 
Note that $\phi$ determines which projective-polarization measurement the POVM most closely corresponds to. 
Unfortunately, $\beta(\theta)$ reaches a maximum of $0.6918$ for $\theta_{\text{opt}}=81.66^{\circ}$, and so perfect Compton polarimetry after a single scattering event is impossible.
The Supplemental Material contains further discussion of this POVM in relation to quantum state fidelity, quantum state discrimination, and the Helstrom bound~\cite{supp}.

Consider now generalizing to $N$ sequential scattering events prior to polarization-insensitive detection by a scintillator. For an initial state with Stokes vector $\ket{S}$, the $N$-fold differential scattering cross section is given by~\cite{mcmaster1961matrix,caradonna2024kinematic} 
\begin{align}\label{eq:N-fold-DSC}
    \dfrac{{\rm d}^N\sigma}{{\rm d}\Omega_1 d\Omega_2\ldots d\Omega_N}&=\bra{I}T_{N}M_{N}\ldots T_{2}M_{2}\,T_{1}M_{1}\,\ket{S},
\end{align}
where $T_{j}$ and $M_{j}$ are the Compton transition and rotation Mueller matrices that describe the $j^{\text{th}}$ scattering event, each evaluated at the photon energy $E_{j-1}$ propagated from the preceding event, and are given explicitly in the End Matter. 
Here, polarization-insensitive detection is described by projection onto $\bra{I}=(1,0,0,0)$, corresponding to the maximally-mixed photon polarization state.
Prior to this projection, $\left(T_{N}M_{N}\ldots T_{2}M_{2}\,T_{1}M_{1}\right)\,\ket{S}$ is a Stokes vector that describes an unnormalized state, which effectively keeps track of the probabilities of sequential scatterings.
Furthermore, the sequential matrix product assumes that each scattering event is independent and there is no quantum coherence between the electrons involved in distinct scattering events. 
This condition holds true here as the distances between scatterers is much greater than the photon's wavelength ($10^{-12}{\rm\, m}$ for a $0.5{\rm\, MeV}$ photon).
Using Eq.~\eqref{eq:N-fold-DSC}, the POVM element corresponding to $N$ scattering events $\Pi^{(N)}(\bm{\theta},\bm{\phi})$ through a trajectory defined by $\bm{\theta}=(\theta_1,\theta_2,\ldots,\theta_N)$ and $\bm{\phi}=(\phi_1,\phi_2,\ldots\phi_N)$ may also be deduced, proceeding as in the single scattering case.

For the purposes of our Bell test, however, we focus on a particular set of \emph{co-planar} trajectories. Namely, we consider trajectories where the first scattering event $(\theta_{1},\phi_{1}=\phi_{\pm})$ is followed by sequential scattering events that only occur in the polar direction $(\theta_j,\phi_j=0)$ $\forall\,j>1$ defining a $2\text{D}$ scattering plane. Consequently, $M(\phi_j)=\mathds{1}_{4}$ for $j>1$, which allows the total, $N$-Compton transition matrix to maintain a block-diagonal structure. Crucially, this in turn implies that $\Pi^{(N)}(\bm{\theta},\phi)$ retains the same form as Eq.~\eqref{eq:POVM1} and that the dichotomic, normalized polarization POVM operators simply read: 
\begin{align}\label{eq:POVMN}
    \overline{\Pi}_{\pm}(\bm{\theta},\phi)=\dfrac{1}{2}\begin{pmatrix}
        1 & \pm\beta(\bm{\theta})e^{-2i\phi} \\
        \pm\beta(\bm{\theta})e^{2i\phi} & 1
    \end{pmatrix},
\end{align}
where we deliberately emphasize the dependence of the measurement on the choice of the {\em first} azimuthal scattering angles, parametrized by $\phi=\phi_{\pm}-\pi/4\pm \pi/4$.
However, whereas for a single scattering event $\beta(\theta)\leq\beta({\theta_{\text{opt}}})=0.6918$, the function $\beta(\bm{\theta})$ corresponding to $N$ sequential scattering events can exceed this value. In this way, the POVM for $N\geq2$ can be closer to an ideal projective measurement of linear polarization. Notice that, in this POVM, the distinction between the two outcomes occurs at the first scattering event, whilst the following trajectories (dictating the positioning of detectors, in practice) are fixed. 
We numerically optimize the multi-variable function $\beta(\bm{\theta})$ over $\bm{\theta}$ to find $\bm{\theta}_{\text{opt}}$ and the results up to $N=10$ are given in Tables~\ref{tab:optimal_angles} and \ref{tab:chsh_results} in the End Matter.
Numerically, in the limit of many scattering events we observe $\beta(\bm{\theta}_{\text{opt}})\rightarrow1$, and so Eq.~\eqref{eq:POVMN} realizes an ideal projective measurement of linear polarization.


%
\textit{Bell Test of Annihilation Photons.}---In entanglement theory, a witness is an observable whose expectation value discriminates a subset of entangled states from the set of all possible separable states~\cite{horodecki2009quantum}.
Although previous works demonstrate consistency with entanglement, they are not entanglement witnesses in QI theory, as they either admit false positives from separable states~\cite{bohm1957discussion,watts2021photon,moskal2025nonmaximal,caradonna2024stokes,bordes2024first} or do not account for the inherent insensitivity of Compton scattering to circular polarization~\cite{hiesmayr2019witnessing}, which is apparent from $\bra{R}\Pi^{(1)}(\theta,\phi)\ket{R}=\bra{L}\Pi^{(1)}(\theta,\phi)\ket{L}=\mathcal{N}$. 
We discuss these issues in detail in the Supplemental Material~\cite{supp}. Crucially, Refs~\cite{kasday1971experimental,kasday1972distribution,pei2026can} demonstrate that witnesses constructed from the Pryce--Ward cross section~\cite{pryce1947angular}---corresponding to $N=1$ scattering event per annihilation photon---admit a LHV description, and so cannot certify entanglement.
However, a Bell test resolves these issues by providing a valid entanglement witness~\cite{terhal2000bell} \emph{and} also ruling out LHV models~\cite{bell1964onthe}.
%

\begin{figure*}
    \centering
    \includegraphics[width=0.9\linewidth]{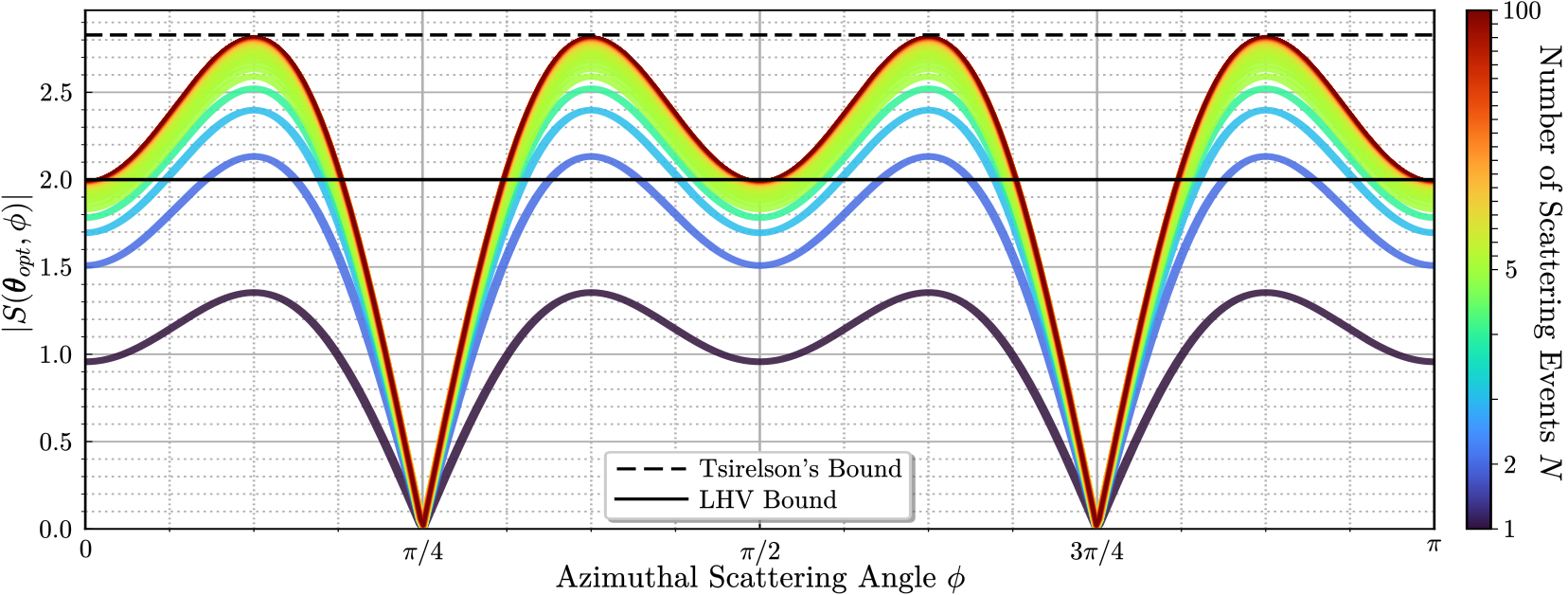}
    \caption{Plot of the CHSH function optimized over polar scattering angles $\lvert S(\bm{\theta}_{\text{opt}},\phi)\rvert$ for $N=1$ to $N=100$ scattering events. Even for $N=2$, the LHV bound $=2$ is exceeded around $\phi=(2m+1)\pi/8$ with $m\in\mathds{Z}$, and as $N$ increases Tsirelson's bound of $2\sqrt{2}$ is approached. Alice and Bob's measurement settings are $\{\phi_a=0,\phi_{a'}=2\phi\}$ and $\{\phi_b=-\phi,\phi_{b'}=-3\phi\}$, respectively.}
    \label{fig:CHSH}
\end{figure*}

The setup for such a Bell test, generalizing the original KUW experiment, is shown in Fig.~\ref{fig:setup}(c).
Following the annihilation of an electron-positron pair at rest, or in the ground state of parapositronium, conservation of parity, angular momentum, and energy demand the two-photon decay channel produces the entangled Bell state $\ket{\Phi^{-}}=\frac{1}{\sqrt{2}}\left(\ket{R}\ket{R}-\ket{L}\ket{L}\right)$ with each photon having initial energy $E_0=1$~\cite{kasday1972distribution}. 
To verify this entanglement, Alice and Bob make local measurements on the photons that comprise this entangled pair. These measurements are realized by $N$ Compton scattering events followed by photon detection at either $\phi_+$ or $\phi_-$, which are assigned the measurement outcomes $+1$ or $-1$, respectively. 
By postselecting on co-planar trajectories corresponding to a given $\bm{\theta}$, each local measurement is completely specified by the azimuthal angle $\phi$ and is described by the POVM with elements $\overline{\Pi}_{\pm}(\bm{\theta},\phi)$ of Eq.~(\ref{eq:POVMN}). In this way, Alice and Bob realize non-projective dichotomic measurements on orthogonal linear polarization states. 

From their measurement statistics, the CHSH function
\begin{align}\label{eq:CHSH_func}
    S=E(\phi_a,\phi_b)-E(\phi_a,\phi_{b'})+E(\phi_{a'},\phi_b)+E(\phi_{a'},\phi_{b'}),
\end{align}
may be calculated, where $\{\phi_a,\phi_{a'}\}$ and $\{\phi_b,\phi_{b'}\}$ represent two possible measurement settings for Alice and Bob, and the expectation values associated with these settings are, for example,
\begin{align}\label{eq:eval_exp}
    E(\phi_a,\phi_b)=\dfrac{N_{++}+N_{--}-N_{+-}-N_{-+}}{N_{++}+N_{--}+N_{+-}+N_{-+}}.
\end{align}
Here, $N_{+-}$ corresponds to the number of coincidence counts where Alice measures the outcome $+1$ and Bob measures the outcome $-1$. For any classical LHV theory, the CHSH inequality states that $\lvert S\rvert\leq2$~\cite{clauser1969proposed}. Quantum mechanics predicts that entangled states may achieve $\lvert S\rvert>2$, with the maximal violation achievable $2\sqrt{2}$ known as Tsirelson's bound~\cite{cirel1980quantum}. In the limit of a large number of measurements, the coincidence counts in Eq.~\eqref{eq:eval_exp} may be replaced by the probabilities calculated via our POVM-based analysis, e.g.,
\begin{align}\label{eq:Prob_phia_phib}
    \frac{N_{+-}}{N_{++}+N_{--}+N_{+-}+N_{-+}}=\bra{\Phi^{-}}
    \overline{\Pi}_{+}(\phi_a)\otimes\overline{\Pi}_{-}(\phi_b)\ket{\Phi^{-}}.
\end{align}

 In general, $S$ depends on 4 azimuthal angles corresponding to each measurement setting Alice and Bob may select. However, if the so-called ``Bell-test angles''~\cite{kasday1975angular,aspect1982experimental}, $\phi_a=0$, $\phi_{a'}=2\phi$, $\phi_b=-\phi$, and $\phi_{b'}=-3\phi$, are chosen, then Eqs~\eqref{eq:POVMN} to \eqref{eq:Prob_phia_phib} evaluate to 
\begin{align}\label{eq:Sfunc_N}
    S(\bm{\theta},\phi)=\beta^2(\bm{\theta})\left(-3\cos2\phi+\cos6\phi\right).
\end{align}
We detail the steps of this derivation in the Supplemental Material~\cite{supp}.

From Eq.~\eqref{eq:Sfunc_N}, the function $S$ is maximized at $\phi=(2m+1)\pi/8$, with $m\in\mathds{Z}$, and when the multi-variable function $\beta(\bm{\theta})$ is also maximized. Again, these optimal angles $\bm{\theta}_{\text{opt}}$, given in Table~\ref{tab:optimal_angles}, correspond to the configuration where $\Pi^{(N)}$ is maximally close to an ideal projective measurement of linear polarization. At $\phi=\pi/8$, Alice's measurement settings are maximally close to measurements in the bases $\{\ket{H},\ket{V}\}$ and $\{\ket{A},\ket{D}\}$, while Bob's measurements represent points on the equator of the Bloch sphere equidistant between each of these eigenstates.

In Fig.~\ref{fig:CHSH}, we plot $\lvert S(\bm{\theta}_{\text{opt}},\phi)\rvert$ for $N=1$ to $100$. For $N=1$, which corresponds to the Pryce--Ward cross section, our results agree with the original KUW experiment and there is no value of $\phi$ for which the LHV bound is violated as $\beta(\theta_{\text{opt}})$ is too low~\cite{kasday1975angular}. However, for $N\geq 2$---going beyond the Pryce--Ward cross section---the function $\beta(\bm{\theta})$ can be large enough to observe a violation of the CHSH inequality, providing a genuine witness for entanglement in this system. Indeed, in the limit of many scattering events, $\beta(\bm{\theta}_{\text{opt}})\rightarrow1$ and Tsirelson's bound is approached. See Table~\ref{tab:chsh_results} in the End Matter, which lists $\beta$ and the maximum $S$ for the optimal angles $\bm{\theta}$ up to $N=10$.

Having established a POVM-based framework that enables a Bell test, and thus a definitive entanglement witness, using annihilation photons, we now briefly consider the experimental feasibility of such a measurement. We have simulated the simplest (i.e., $N = 2$) experiment capable of violating a CHSH inequality using the \textsc{Geant4} code~\cite{collaboration2003geant4}. This simulation, which is described in more detail in the Supplemental Material~\cite{supp}, gives a ratio of the total rate of coincidence counts, $\mathcal{C}$, to the rate of photon pair production, $r$, equal to $\mathcal{C}/r=5\times10^{-13}$. A complementary analytic calculation via Eq.~\eqref{eq:Prob_phia_phib} which accounts for entanglement, and uses scattering probabilities from the \textsc{Geant4} simulation, is consistent with this result. The obtainable experimental rate will depend on the available source activity. Realistically, the maximum available source intensity would be on the order of $1\,\text{GBq}$, yielding $\mathcal{C}\approx 300$ coincidence counts per week. In order to establish the number of detected events and associated SNR required to establish a statistically valid Bell test using this method a dedicated feasibility study is required, which is currently ongoing, and will be presented in future work: our preliminary indications are that it is likely that, even under optimal conditions, measurement times of several months would be required.  


\textit{Conclusion and Outlook.}---By describing Compton polarimetry using the tools of quantum information theory, and generalizing to $N$ sequential scattering events, we have predicted Bell inequality violations by the entangled photons produced from electron-positron annihilation for $N\geq2$.
The POVM-based description of Compton scattering we introduce here is applicable to contexts well beyond this study, such as quantum-enhanced PET scanners~\cite{watts2021photon}, high-precision Compton telescopes for gamma-ray astronomy~\cite{yoneda2023reconstruction} and inverse-Compton scattering constraints to dark-matter models \cite{cirelli2009inverse}.
Furthermore, our POVM-based analysis can be extended to investigate a wide range of scattering processes from a quantum-information perspective, including other QED processes~\cite{cervera2017maximal,fedida2023tree}, weak interactions~\cite{cervera2019maximal,ashby2023quantum}, and QCD processes~\cite{liu2023minimal,barr2024quantum,nunez2026universality}. 

At the heart of Wheeler's original proposal was the notion that photon-polarization correlations produced from electron-positron annihilation could be used to investigate the surprising nature of quantum entanglement~\cite{wheeler1946polyelectrons}. 
Although it is impossible to violate a Bell inequality via the original KUW experiment~\cite{kasday1971experimental,kasday1972distribution,kasday1975angular}, here we have demonstrated that it is in fact possible to certify entanglement in the two-photon state produced by electron-positron annihilation. 
By going beyond the Pryce--Ward cross section, our proposal therefore provides a clear experimental path towards an unambiguous Bell-inequality violation and a route for further studies of entanglement at the MeV scale. 
At even higher energies, analogous issues arise in collider physics. Recent critical analyses demonstrate that the relevant differential cross sections admit an LHV construction~\cite{abel1992testing,abel2025colliders,bechtle2025critical}, placing the entanglement claims of Refs~\cite{atlas2024observation,cms2024observation} on similar footing to the $N=1$ case we discuss. Our results suggest that genuine entanglement certification in colliders may also require measurement strategies that go beyond single scattering events.


\begin{acknowledgments}
\textit{Acknowledgements.}---We acknowledge useful discussions
with Donovan M. Newson, Andreas Lanz, and Daniel P. Watts. JC and AS acknowledge funding from the Leverhulme Trust through Research Project Grant No. RPG-2024-287. JW and DBC acknowledge funding from the Leverhulme Trust through Research Project Grant No. RPG-2024-184.
\end{acknowledgments}



\bibliography{bib1}


\section{End Matter}
\subsection{The functions $\alpha$ and $\beta$}
The $\alpha$ and $\beta$ functions appearing in Eq.~\eqref{eq:KN_1a}, corresponding to a single scattering event, are given by
\begin{align}
    \alpha(E_{0},\theta)&=\frac{1}{2}\left(\frac{E}{E_0}\right)^2\left({\frac{E}{E_{0}}+\frac{E_{0}}{E}-\sin^2\theta}\right)\\
    \beta(E_{0},\theta)&=\dfrac{\sin^2\theta}{\frac{E}{E_{0}}+\frac{E_{0}}{E}-\sin^2\theta},
\end{align}
where the scattered photon energy is
\begin{align}
    E&=\dfrac{E_{0}}{1+E_{0}\left(1-\cos\theta\right)}.
\end{align}
At initial energy $E_0=1$, relevant for electron-positron annihilation at rest, these functions reduce to
\begin{align}
    \alpha(E_{0}=1,\theta)=\frac{\cos \theta \left[ 3 + (\cos \theta - 3) \cos \theta \right]-3}{2 (\cos \theta - 2)^3},\\
    \beta(E_{0}=1,\theta)=\frac{ (\cos \theta - 2) \sin^2 \theta}{\cos \theta [3 + (\cos \theta - 3) \cos \theta] - 3}.
\end{align}
The functions $\alpha(\bm{\theta})$ and $\beta(\bm{\theta})$, corresponding to $N\geq2$ scattering events, are discussed in the Supplemental Material~\cite{supp}. 

\subsection{Stokes--Mueller Description of Compton Scattering}
The Compton transition matrix and Mueller rotation matrix for the $j^{\text{th}}$ scattering event are ~\cite{mcmaster1961matrix,caradonna2024kinematic}
\begin{align}
    T_{j}&=\frac{r_e^2}{2}\left(\dfrac{E_{j}}{E_{j-1}}\right)^2
    \begin{pmatrix}
       t_{11}& t_{12} & 0 & 0 \\
       t_{12} & 2-t_{12} & 0 & 0 \\
       0 & 0 & t_{33} & 0 \\
       0 & 0 & 0 & t_{44}
    \end{pmatrix},\\
    M_{j}&=\begin{pmatrix}
    1 & 0 & 0 & 0 \\ 
    0 & \cos(2\phi_j) & \sin(2\phi_j) & 0 \\
    \\ 0 & -\sin(2\phi_j) & \cos(2\phi_j) & 0 \\
    0 & 0 & 0 & 1
\end{pmatrix},
\end{align}
where
\begin{align}
    E_j&=\dfrac{E_{j-1}}{1+E_{j-1}\left(1-\cos\theta_j\right)},\nonumber\\
    t_{11}(E_{j-1},\theta_j)&=1+\cos^2\theta_j+\left(E_{j-1}-E_j\right)\left(1-\cos\theta_j\right),\nonumber\\
    t_{12}(\theta_j)&=\sin^2\theta_j, \nonumber\\
    t_{33}(\theta_j)&=2\cos\theta_j,\nonumber\\
    t_{44}(E_{j-1},\theta_j)&=2\cos\theta_j+\left(E_{j-1}-E_j\right)\left(1-\cos\theta_j\right)\cos\theta_j.
\end{align}
In a single Compton scattering event, a photon with initial Stokes vector $\ket{S}$ with energy $E_0$ interacts with a stationary electron in the maximally mixed spin state. 
Straight out of Compton scattering, the photon propagates in a direction defined by angles $(\theta,\phi)$ from its initial momentum direction, and the electron degrees of freedom are traced out. In the polarization subspace of the photon, this is described by the transformation $\ket{S}\rightarrow T(E_{0},\theta)M(\phi)\ket{S}$. For sequential scatterings, the evolution of the Stokes vector generalizes to $\ket{S}\rightarrow T_{N}M_{N}\ldots T_{2}M_{2}\,T_{1}M_{1}\,\ket{S}$. Note that after scattering, the Stokes vector describes an unnormalized state, i.e. the element of this vector at position $S_0$ is not equal to 1. As polarization-insensitive detection is described by projection onto $\bra{I}$, the $N$-fold differential cross section is therefore given by Eq.~\eqref{eq:N-fold-DSC}, which projects out the $S_0$ element. In this way, the Stokes vector keeps track of the probabilities of sequential scatterings.

\subsection{Optimal Polar Scattering Angles and Results}
\begin{table}[h!]
\caption{Optimal Scattering Angles $\bm{\theta}_{\text{opt}}=(\theta_1,\theta_2,\ldots,\theta_N)$ in radians up to $N=10$.}
\label{tab:optimal_angles}
\begin{tabular}{lcccccccccc}
\toprule
$N$ & $\theta_{1}$ & $\theta_{2}$ & $\theta_{3}$ & $\theta_{4}$ & $\theta_{5}$ & $\theta_{6}$ & $\theta_{7}$ & $\theta_{8}$ & $\theta_{9}$ & $\theta_{10}$ \\
\midrule
1 & 1.425 &  &  &  &  &  &  &  &  &  \\
2 & 1.038 & 1.479 &  &  &  &  &  &  &  &  \\
3 & 0.777 & 1.071 & 1.499 &  &  &  &  &  &  &  \\
4 & 0.641 & 0.794 & 1.089 & 1.510 &  &  &  &  &  &  \\
5 & 0.557 & 0.653 & 0.805 & 1.100 & 1.517 &  &  &  &  &  \\
6 & 0.498 & 0.565 & 0.661 & 0.813 & 1.109 & 1.521 &  &  &  &  \\
7 & 0.454 & 0.505 & 0.571 & 0.667 & 0.819 & 1.115 & 1.525 &  &  &  \\
8 & 0.420 & 0.460 & 0.510 & 0.576 & 0.672 & 0.823 & 1.120 & 1.528 &  &  \\
9 & 0.393 & 0.425 & 0.464 & 0.514 & 0.580 & 0.676 & 0.827 & 1.124 & 1.530 &  \\
10 & 0.370 & 0.397 & 0.428 & 0.467 & 0.517 & 0.584 & 0.679 & 0.830 & 1.127 & 1.532 \\
\bottomrule
\end{tabular}
\end{table}

\begin{table}[h!]
\caption{Values of $\beta$, $\mathrm{max}(|S|)$, fidelity $F^{(N)}$, trace-norm distance $ D^{(N)}$, and final photon energy $E_N$ at the optimal angles $\bm{\theta}_{\text{opt}}$ up to $N=10$. The values for $N=50$ and $N=100$ are also shown here.}
\label{tab:chsh_results}
\begin{tabular}{cccccc}
\toprule
$N$ & $\beta$ & $\mathrm{max}(|S|)$ & $ F^{(N)}$ & $ D^{(N)}$ & $E_N$ \\
\midrule
1 & 0.6918 & 1.3537 & 0.8459 & 0.1541 & 0.5391 \\
2 & 0.8683 & 2.1326 & 0.9342 & 0.0658 & 0.4166 \\
3 & 0.9207 & 2.3977 & 0.9604 & 0.0396 & 0.3655 \\
4 & 0.9440 & 2.5205 & 0.9720 & 0.0280 & 0.3363 \\
5 & 0.9569 & 2.5900 & 0.9785 & 0.0215 & 0.3168 \\
6 & 0.9651 & 2.6343 & 0.9825 & 0.0175 & 0.3026 \\
7 & 0.9707 & 2.6649 & 0.9853 & 0.0147 & 0.2915 \\
8 & 0.9747 & 2.6873 & 0.9874 & 0.0126 & 0.2826 \\
9 & 0.9778 & 2.7043 & 0.9889 & 0.0111 & 0.2752 \\
10 & 0.9802 & 2.7177 & 0.9901 & 0.0099 & 0.2689 \\
50 & 0.9964 & 2.8078 & 0.9982 & 0.0018 & 0.1991 \\
100 & 0.9982 & 2.8182 & 0.9991 & 0.0009 & 0.1791 \\
\bottomrule
\end{tabular}
\end{table}

In Table~\ref{tab:optimal_angles}, for $N=1$, $\theta_{1}=\theta_{\text{opt}}=1.425\,\text{rad}=81.66^{\circ}$. To violate a CHSH inequality, Eq.~\eqref{eq:Sfunc_N} implies that $\beta>2^{-1/4}\approx0.8409$, which is satisfied for $N\geq2$.
In Table~\ref{tab:chsh_results}, the final energy $E_N$ refers to the energy of the photon impinging on the detector after its last scattering events in units of electron rest mass.
The quantum state fidelity $F^{(N)}$ and the trace-norm distance $ D^{(N)}$, quantify how close the POVM is to an ideal projective polarization measurement, and are discussed in detail in the Supplemental Material~\cite{supp}.
For $N=50$, $\bm{\theta}_{\text{opt}}=\left(0.1569,0.1588,\ldots,1.5496\right)\,\text{rad}$ and for $N=100$, $\bm{\theta}_{\text{opt}}=\left(0.110,0.111,\ldots,1.556\right)\,\text{rad}$.
%


\clearpage
\onecolumngrid
\setcounter{equation}{0}
\setcounter{figure}{0}
\setcounter{table}{0}
\setcounter{page}{1}
\makeatletter
\renewcommand{\theequation}{S\arabic{equation}}
\renewcommand{\thefigure}{S\arabic{figure}}
\renewcommand{\thesection}{\arabic{section}} 
\makeatother
\setcounter{secnumdepth}{3}

\begin{center}
    {\large \textbf{Supplemental Material: \\ Bell Test of Photons from Electron-Positron Annihilation \\ via POVM-based Compton Polarimetry}}
\end{center}
\vspace{20pt}

\begin{centering}
Jack~Clarke\,\orcidlink{0000-0001-8055-449X},$^{1}$~  
Preslav Asenov\,\orcidlink{0009-0003-4232-4521},$^{1}$~
Jesse Smeets\,\orcidlink{0009-0009-5205-4936},$^{2,\,3}$~ \\
Jia-Shian Wang\,\orcidlink{0000-0003-2499-7039},$^{1}$~ 
David B. Cassidy\,\orcidlink{0000-0001-8332-5553},$^{1}$~
and Alessio Serafini\,\orcidlink{0000-0002-4509-7470}$^{1}$\\
\end{centering}

\vspace{16pt}

\begin{centering}
\textit{\small
$^1$ Department of Physics and Astronomy, University College London, Gower Street, WC1E 6BT London, United Kingdom \\
$^2$ Department of Physics and Astronomy, LaserLaB, VU Amsterdam, De Boelelaan 1100, Amsterdam, 1081 HZ, The Netherlands \\
$^3$ Department of Applied Physics, Eindhoven University of Technology, 513, 5600 MB Eindhoven, The Netherlands \\
}
\end{centering}

%
\begin{quote}
{\small{In this Supplemental, we first discuss previous proposals for witnessing polarization entanglement in the two-photon state produced by electron-positron annihilation. 
We clarify the quantum-information-theoretic requirements for a valid entanglement witness and discuss the relationship between our approach and these existing methods. In particular, we describe how these previous works do not constitute entanglement witnesses, while our Bell test is strictly stronger, not only verifying entanglement, but also ruling out local hidden variable theories.
Following this, we provide further details on our POVM-based analysis of Compton polarimetry and generalization to $N$ scattering events. We then provide further details on the derivation of the CHSH function for the generalized-KUW experiment. Finally, we discuss the \textsc{Geant4} simulations for the $N=2$ experiment.}}
\end{quote}


\section{Previous Entanglement Claims}
\subsection{Entanglement Witnesses and Nonlocality}
For a bipartite entangled state $\rho$, an entanglement witness $W$ can be defined as an Hermitian operator that satisfies $\Tr\left(\rho W\right)>0$ and $\Tr\left(\omega W\right)\leq0$ for all separable states $\omega$~\cite{horodecki2009quantum}. We note the value of zero here is a matter of convention, and the operator $W$ can be shifted by a constant multiplied by the identity operator to shift the hyperplane that separates the entangled state $\rho$ from all separable states. Further, note that a negative constant will flip the sign of the above inequalities. Nevertheless, if the above condition is not satisfied by some observable $W$, then it does not provide a genuine entanglement witness.

In the context of our generalized-KUW experiment, the entanglement witness is $W=\mathcal{B}-2$, where $\mathcal{B}$ is the Bell operator appearing in the CHSH inequality~\cite{terhal2000bell}. Note that this witness may be considered to be non-optimal~\cite{hyllus2005relations}, but it is operational in the sense that it can be measured through the generalized-KUW test we describe in the main text. Furthermore, in addition to being an entanglement witness, a Bell test rules out all local hidden variable (LHV) models, so is strictly stronger than an entanglement witness.

\subsection{False Positive From Separable States and Local Hidden Variable Theories}
The observables studied in previous works were developed primarily within the high-energy physics community, where the term ``entanglement witness'' is used differently compared to its precise quantum-information-theoretic meaning. Here, we clarify this distinction, which is essential for establishing an unambiguous proof of entanglement. In particular, we demonstrate that these observables admit false positives from separable states. Although these metrics can discriminate between predicted entangled states and particular separable alternatives, they cannot objectively rule out the entire set of separable states as a formal Bell test would. Moreover, we discuss how the observables studied in previous works cannot rule out LHV models as they are constructed from the Pryce--Ward cross section~\cite{pryce1947angular}, which is known to admit a LHV description~\cite{kasday1971experimental,kasday1972distribution}.

\subsubsection{The R Ratio}
In the 2-photon decay channel, the double differential scattering cross section of the back-to-back Compton-scattered photons ${d^4\sigma}/{d\Omega_{a}\Omega_{b}}$ (the Pryce--Ward cross section~\cite{pryce1947angular}) depends on $\theta_{a}$, $\theta_b$, and the difference between azimuthal angles $\Delta \phi_{ab}$ of the two photons. 
The ratio of counts defined by
\begin{align}
    R&=\max_{\left\{\theta_a,\theta_b,\Delta \phi,\pm\right\}}\dfrac{\dfrac{d^4\sigma}{d\Omega_{a}d\Omega_{b}}(\theta_a,\theta_b,\Delta \phi_{ab}=\pm90^{\circ})}{\dfrac{d^4\sigma}{d\Omega_{a}d\Omega_{b}}(\theta_a,\theta_b,\Delta \phi_{ab}=0^{\circ})},
\end{align}
is referred to as an ``entanglement witness'' by many sources---see for example Refs~\cite{watts2021photon,caradonna2024stokes,bordes2024first,moskal2025nonmaximal} cited within the main. Namely, if $R>1.63$, the claim is that entanglement is witnessed. Indeed, $R$ finds its origin in Ref.~\cite{bohm1957discussion}, which was published many decades before the term \emph{entanglement witness} was introduced in quantum information theory~\cite{terhal2000bell}. However, there exist separable states $\omega$ that also give $R>1.63$. 

Firstly, concrete examples of separable states with $E_0=1$ that violate the $R$-ratio bound include the classical mixture $\omega_{\text{mix}}=\frac{1}{2}\left(\ket{HV}\bra{HV}+\ket{VH}\bra{VH}\right)$ ($R=2.84$) and the product state $\ket{HV}$ ($R=5.49$). Thus, $R$ cannot be an entanglement witness. One may dispute this because these states do not respect the expected polarization-symmetries of the annihilation photons. However, this is making an assumption about the underlying physical model and state. In the construction of the ratio $R$, one makes further assumptions about the particular form of the Compton scattering interaction, the structure of the states involved, and, perhaps most profoundly, the applicability of quantum mechanics at the MeV scale. However, a Bell test makes no such assumptions in its construction, and this is precisely what makes it so powerful.

Secondly, consider the rotationally-symmetric separable state discussed in Ref.~\cite{caradonna2024stokes}
\begin{align}
    \omega_{\text{sep}}=\frac{1}{4}\begin{pmatrix}
        1 & 0 & 0 & -1 \\
        0 & 1 & 0 & 0 \\
        0 & 0 & 1 & 0 \\
        -1 & 0 & 0 & 1
    \end{pmatrix},
\end{align}
which do respect the expected polarization-symmetries of the annihilation photons. For $E_0=1$, $\omega_{\text{sep}}$ gives $R=1.63$. However, for any energy lower than $E_{0}=1$, $R>1.63$ and in the limit $E_{0}\rightarrow0$, $R=2.7$. Thus, $R$ cannot be a model-independent certification of entanglement. A valid model-independent entanglement witness must, by definition, certify entanglement independently of assumptions about the physical model generating the state. An observable that requires prior knowledge of $E_0$ to correctly interpret its outcome is therefore not a model-independent witness, which is precisely the operational power that a Bell inequality provides.

The fact that $R$ cannot constitute an unambiguous proof of entanglement is consistent with the motivation of the KUW experiment itself, where the authors recognized that a Bell inequality test---rather than a ratio-based observable~\cite{wu1950theangular}---was the appropriate standard for witnessing entanglement~\cite{kasday1972distribution,kasday1975angular}. We believe the recent use of the term ``entanglement witness'' in this context is therefore simply a misnomer arising from the fields of quantum information and high-energy physics colliding at high velocities.

\subsubsection{The Normalized Correlation Observable and Local-Hidden-Variable Models}
In Ref.~\cite{pei2026can} the authors introduce the normalized correlation observables $\mathcal{O}_{1}$. They claim that the predicted range for entangled states is $\expval{\mathcal{O}_{1}}\in[-1,1]$ and the predicted range for any separable state $\omega$ is $\expval{\mathcal{O}_{1}}\in[-1/2,1/2]$. Thus, they state that any deviation from $[-1/2,1/2]$ is sufficient evidence for quantum entanglement. However, the authors themselves demonstrate that, without imposing an additional mirror symmetry constraint, a LHV theory can reproduce the full joint angular distribution of the final photon state, including values of $\expval{\mathcal{O}_{1}}$ outside of $[-1/2,1/2]$. Therefore, a deviation of $\expval{\mathcal{O}_{1}}$ beyond this range cannot constitute model-independent evidence for entanglement, since a local realistic description remains viable unless mirror symmetry is separately assumed. As with the ratio $R$, the certification of entanglement depends on assumptions about the underlying physical model, which is precisely what a Bell inequality test avoids.
Furthermore, as with the ratio $R$, for $E_0<1$, values of $\expval{\mathcal{O}_{1}}$ outside of $[-1/2,1/2]$ can be found for separable states. As a concrete example, consider the state $\omega_{\text{mix}}$. Indeed, at $E_0=1$, $\expval{\mathcal{O}_{1}}=-1/2$, but for any $E_0<1$ the separable state $\omega_{\text{mix}}$ gives $\expval{\mathcal{O}_{1}}<-1/2$. Moreover, as $E_0$ decreases, $\expval{\mathcal{O}_{1}}$ can deviate even further than the range $[-1,1]$ predicted for entangled states. 

Ref.~\cite{pei2026can} constitutes a detailed examination of the role of mirror symmetry in entanglement in this system, complementing previous notable studies in this direction~\cite{caradonna2024stokes}. Furthermore, their results confirm earlier results by Kasday~\cite{kasday1971experimental,kasday1972distribution}---see, for example, Appendix C of Kasday's thesis---namely, that LHV theories can reproduce the Pryce--Ward cross section. Therefore, an observable built purely from the Pryce--Ward cross section cannot produce a genuine certification of entanglement, as it does not contain enough information about the polarization degree of freedom. This is why our Bell test needs $N\geq2$ scattering events to be successful, going beyond the limits of the Pryce--Ward cross section ($N=1$).
Notably, the ``entanglement witnesses'' discussed elsewhere in this section are built from the Pryce--Ward cross section. 

\subsection{Witness via Mutually Unbiased Bases}
Ref.~\cite{hiesmayr2019witnessing} outlines a strategy to witness entanglement in electron-positron annihilation. Here, the authors utilize a Kraus-type operator approach and explore its formal consequences, but unlike our approach, these operators do not correspond to the outcomes of momentum (or equivalently, $\theta$ and $\phi$) measurements. We would like to briefly clarify that, although there has been recent discourse in the literature regarding certain results of Ref.~\cite{hiesmayr2019witnessing}---see, for example, Refs~\cite{caradonna2019probing,ivashkin2023testing,caradonna2024stokes,zugec2026reconciliation}---our following discussion is separate from this prior discourse and focuses specifically on the claim regarding the witnessing of entanglement.
To summarize, the strategy outlined in Ref.~\cite{hiesmayr2019witnessing} cannot witness entanglement as a standard Compton polarimeter is blind to circular polarization. 

For qubits, the three bases $\{\ket{H},\ket{V}\}$, $\{\ket{A},\ket{D}\}$, and $\{\ket{R},\ket{L}\}$, corresponding to eigenvectors of the Pauli matrices, are mutually unbiased bases (MUBs). Following Ref.~\cite{hiesmayr2019witnessing}, for $m$ MUBs, the inequality
\begin{align}
    \dfrac{m-1}{2}\leq I_{m}(\omega) \leq 1+\dfrac{m-1}{2}
\end{align}
holds for separable states $\omega$, where for a general state $\rho$
\begin{align}
    I_{m}(\rho)&=\sum_{k=1}^{m}\sum_{l=0}^{2} \text{Prob}(i_{l}^{(k)},j_{l}^{(k)}).
\end{align}
Here, the index $m$ denotes the MUB, the index $l$ spans the two eigenvectors within an MUB, $i_{l}^{(k)}$ and $j_{l}^{(k)}$ are the eigenvalues of eigenvectors that form the MUBs, and $\text{Prob}(i_{l}^{(k)},j_{l}^{(k)})$ refers to a joint probability corresponding to a measurement of the state $\rho$ in the chosen MUB. For the relevant case of two-dimensional systems, the maximum number of MUBs is $m=3$, and therefore the authors of Ref.~\cite{hiesmayr2019witnessing} claim that any deviation from $I_{3}\in[1,2]$ witnesses entanglement. Here, we now focus on the implicit claim that one has access to the $\{\ket{R},\ket{L}\}$ MUB in standard Compton scattering.

In Compton polarimetry, as considered in Ref.~\cite{hiesmayr2019witnessing}, the scattering distribution reveals information about the degree of linear polarization, but is completely insensitive to circular polarization~\cite{del2024compton}. In the Klein--Nishina formula for polarized photons, this is evident from the dependence on the Stokes parameters $S_{1}$ and $S_{2}$, and the absence of the Stokes parameter $S_{3}$, which corresponds to circular polarization---see Eq.~\eqref{eq:KN_1} below. Therefore, the number of MUBs one has access to is $m=2$. Thus, the authors of Ref.~\cite{hiesmayr2019witnessing} should instead consider $I_{2}$ and claim any deviation from $I_{2}\in[0.5,1.5]$ witnesses entanglement. However---using notation from Ref.~\cite{hiesmayr2019witnessing}---for the entangled polarization state $\text{max}\,I_{2}=\frac{1}{2}\left[2+2\mathcal{V}(1,81.67^{\circ})\mathcal{V}(1,-81.67^{\circ})\right]=1.47859<1.5$. Thus, the entangled state cannot violate the $m=2$ separability bound, and is therefore indistinguishable from separable states by this measure since only two MUBs are accessible.
One may use the filtered POVM elements $\overline{\Pi}_{H}$, $\overline{\Pi}_{V}$, $\overline{\Pi}_{D}$, and $\overline{\Pi}_{A}$ (for $N=1$) to confirm this result. Finally, we would like to emphasize that we have not discussed here the underlying physical assumptions that go into the evaluation of $I_{3}$, as we did in our discussion of the ratio $R$.


\section{Further Details on the POVM Analysis}

\subsection{Compton Polarimetry via the Stokes--Mueller Formalism}
\subsubsection{Stokes--Mueller Formalism}
In the Stokes--Mueller formalism, a single Compton scattering event transforms the initial Stokes vector $\ket{S}$ to the Stokes vector $T(E_{0},\theta)M(\phi)\ket{S}$ that describes an unnormalized state~\cite{fano1949remarks,mcmaster1961matrix}. The differential scattering cross section is given by
\begin{align}\label{eq:DSC_1}
    \dfrac{d\sigma}{d\Omega}&=\bra{I}T(E_{0},\theta)M(\phi)\ket{S}.
\end{align}
An arbitrary polarization state $\rho$ is described by the Stokes vector $\ket{S}=(S_0,S_{1},S_{2},S_{3})^{\sf T}$. In the circular basis, this state is
\begin{align}\label{eq:rho_eq_supp}
        \rho=\begin{pmatrix}
        \rho_{RR} & \rho_{RL} \\ \rho_{LR} & \rho_{LL}
    \end{pmatrix}
    =\frac{1}{2}\begin{pmatrix}
        S_{0}+S_{3} & S_{1}-iS_{2} \\ S_{1}+iS_{2} & S_{0}-S_{3}
    \end{pmatrix},
\end{align}
where $S_0=\rho_{RR}+\rho_{LL}$ (equal to $1$ for normalized states; as is the case for the state prior to scattering), $S_1=\rho_{RL}+\rho_{LR}$, $S_2=i\rho_{RL}-i\rho_{LR}$, and $S_3=\rho_{RR}-\rho_{LL}$. 
Inserting Eq.~\eqref{eq:rho_eq_supp} with $S_0=1$ into Eq.~\eqref{eq:DSC_1} and evaluating the matrix product gives the Klein--Nishina formula for an arbitrarily polarized photon:
\begin{align}\label{eq:KN_1}
    \frac{d\sigma}{d\Omega} &= \frac{1}{2} r_e^2 \left(\dfrac{E}{E_0}\right)^2 \left[\dfrac{E}{E_0}+\dfrac{E_0}{E}-\sin^2\theta + \sin^2 \theta \left( S_1 \cos 2\phi + S_2 \sin 2\phi \right)\right]\nonumber\\
    &=r_e^2 \alpha(E_{0},\theta) \left[1 + \beta(E_{0},\theta)\left( S_1 \cos 2\phi + S_2 \sin 2\phi \right)\right],
\end{align}
where the $\alpha$ and $\beta$ functions are defined in the End Matter. When $S_{1}=S_{2}=0$, one recovers the standard Klein--Nishina formula for unpolarized light~\cite{klein1928scattering}. Integrating Eq.~\eqref{eq:KN_1} over scattering angles gives the total cross section for one Compton scattering $\sigma_{\text{tot}}^{(1)}$, and normalizing Eq.~\eqref{eq:KN_1} by this quantity gives the scattering probability density ${\rm d}p(\theta,\phi)$ in Eq.~\eqref{eq:KN_1a}. Note that $\sigma_{\text{tot}}^{(1)}=2\pi\int_{0}^{\pi}\,\alpha(\theta)\sin\theta d\theta$, which at $E_0=1$ is $\pi \,r_e^{2}\left[{40}/{9} - 3 \ln(3)\right]$.

\subsubsection{Compton Polarimetry}
A Compton polarimeter determines the degree of linear polarization by comparing count rates at azimuthal angles differing by $90^{\circ}$. In evaluating this, we define the physical azimuthal angle $\phi$ relative to the $x$-axis of the Bloch sphere, so that $\phi = 0$ corresponds to scattering preferentially perpendicular to the plane of horizontal polarization.
 Using this principle, a Compton polarimeter can be used to calculate the ratio of counts
\begin{align}\label{eq:ratio_compton_pol}
   \dfrac{N_{\perp}-N_{\parallel}}{N_{\perp}+N_{\parallel}}&=\frac{\frac{d\sigma}{d\Omega}(\theta,\phi)-\frac{d\sigma}{d\Omega}(\theta,\phi+90^{\circ})}{\frac{d\sigma}{d\Omega}(\theta,\phi)+\frac{d\sigma}{d\Omega}(\theta,\phi+90^{\circ})}=\beta(E_{0},\theta)\left(S_{1}\cos2\phi+S_{2}\sin2\phi\right),
\end{align}
where we have identified $\beta$ as the Compton analyzing power
\begin{align}
    \beta(E_{0},\theta)&=\dfrac{t_{12}(\theta)}{t_{11}(E_{0},\theta)}=\dfrac{\sin^2\theta}{\frac{E}{E_{0}}+\frac{E_{0}}{E}-\sin^2\theta}.
\end{align}
For our Bloch sphere orientation, setting $\phi=0^{\circ}$ corresponds to determining the degree of polarization in the horizontal/vertical basis ($S_{1}$) and $\phi=45^{\circ}$ corresponds to determining the degree of polarization in the diagonal/anti-diagonal basis ($S_{2}$).

\subsection{POVM-based Compton Polarimetry}

\subsubsection{Deducing the POVM Elements}\label{sec:supp_deducing}
Here, we detail how Eq.~\eqref{eq:POVM1} is deduced from Eqs~\eqref{eq:KN_1a} and \eqref{eq:dprob}. Inserting Eq.~\eqref{eq:KN_1a} into the left-hand side of Eq.~\eqref{eq:dprob} gives
\begin{align}\label{eq:deducing}
    \mathcal{N}\left[1+ \beta\left( S_1 \cos 2\phi + S_2 \sin 2\phi \right)\right]={\frac{1}{2}}\sum_{j}S_j {\rm Tr}[\sigma_j \Pi^{(1)}(\theta,\phi)],
\end{align}
which is valid for all $S_1$, $S_2$, $S_3$ and with $S_{0}=1$. Writing the POVM element $\Pi^{(1)}(\theta,\phi)$ in the circular basis
\begin{align}
    \Pi^{(1)}(\theta,\phi)=\begin{pmatrix}
        \Pi_{RR} & \Pi_{RL} \\ 
        \Pi_{LR} & \Pi_{LL}
    \end{pmatrix}
\end{align}
enables the matrix elements of $\Pi^{(1)}(\theta,\phi)$ to be deduced by choosing different values for $S_1$, $S_2$, and $S_3$. For example, when $S_1=S_2=S_3=0$, Eq.~\eqref{eq:deducing} implies
\begin{align}
    {\frac{1}{2}}\left(\Pi_{RR}+\Pi_{LL}\right)=\mathcal{N}.
\end{align}
Moreover, when $S_1=S_2=0$ and $S_{3}\neq0$, 
\begin{align}
    {\frac{1}{2}}\left(\Pi_{RR}-\Pi_{LL}\right)=0.
\end{align}
Also, when $S_2=S_3=0$ and $S_{1}\neq0$,
\begin{align}
    {\frac{1}{2}}\left(\Pi_{LR}+\Pi_{RL}\right)=\mathcal{N}\beta\cos2\phi.
\end{align}
Finally, when $S_1=S_3=0$ and $S_{2}\neq0$,
\begin{align}
    {\frac{1}{2}}\left(\Pi_{LR}-\Pi_{RL}\right)=i\mathcal{N}\beta\sin2\phi.
\end{align}
Solving this system of equations for the elements of $\Pi^{(1)}(\theta,\phi)$ gives
\begin{align}
        \Pi^{(1)}(\theta,\phi)=\mathcal{N}\begin{pmatrix}
        1 & \beta(\theta)e^{-2i\phi} \\
        \beta(\theta)e^{2i\phi} & 1
    \end{pmatrix},
\end{align}
 which is precisely Eq.~\eqref{eq:POVM1} from the main text. Note that as $\int\,\mathcal{N}\,d\Omega =\mathds{1}$, the POVM satisfies $\int\,\Pi^{(1)}(\theta,\phi)\,d\Omega =\mathds{1}$.

\subsubsection{Filtering of POVMs}\label{sec:supp_filtering}

Consider now a general POVM, which is a set of positive operators $\{\Pi_{\mu}\}$ that act on a $d$-dimensional Hilbert space, such that $\sum_{\mu=0}^{n}\Pi_{\mu}={\mathds{1}}$.
Assume the outcomes of this POVM are post-selected restricting to only two values $\mu=0$ and $\mu=1$. Then, for an arbitrary initial state $\rho$, the probabilities $p_0$ and $p_1$ read
\begin{equation}
p_{0} = \frac{{\rm Tr}[\rho \Pi_0]}{{\rm Tr}[\rho \Pi_0]+{\rm Tr}[\rho \Pi_1]} , \quad p_{1} = \frac{{\rm Tr}[\rho \Pi_1]}{{\rm Tr}[\rho \Pi_0]+{\rm Tr}[\rho \Pi_1]},
\end{equation}
such that these post-selected probabilities sum to one. Note that these expressions are in general not even linear in $\rho$. However, for a $d$-dimensional system, if $\Pi_0+\Pi_1 = \frac{1}{d}{\rm Tr}[\Pi_0+\Pi_1]{\mathds{1}}$, then the denominators above simplify to $\frac{1}{d}{\rm Tr}[\Pi_0+\Pi_1]$ and an effective filtered POVM is promptly recovered as
\begin{equation}\label{eq:filtered_POVM_method}
\bar\Pi_0 = \frac{\Pi_0}{\frac{1}{d}{\rm Tr}[\Pi_0+\Pi_1]},
\quad \bar\Pi_1 = \frac{\Pi_1}{{\frac{1}{d}\rm Tr}[\Pi_0+\Pi_1]},
\end{equation}
such that $\overline{\Pi}_{0}+\overline{\Pi}_{1} = {\mathds{1}}$, $p_{0} = {\rm Tr}[\rho \overline{\Pi}_0]$, and $p_{1} = {\rm Tr}[\rho \overline{\Pi}_1]$.

\subsubsection{Compton Polarimetry via Filtered POVMs}

In this work we focus on the POVM elements $\Pi^{(1)}(\theta,\phi)$ described by Eq.~\eqref{eq:POVM1}, which act on a Hilbert space of dimension $d=2$. Post-selecting on the two outcomes, labelled by $\pm$ and determined by the two azimuthal angles $\phi_{+}$ and $\phi_{-}$, gives the two filtered POVM elements via Eq.~\eqref{eq:filtered_POVM_method} as
\begin{align}
    \overline{\Pi}_{\pm}(\theta,\phi)=\frac{1}{2}\begin{pmatrix}
        1 & \pm\beta(\theta)e^{-2i\phi}  \\
        \pm \beta(\theta)e^{2i\phi}  & 1
    \end{pmatrix},
\end{align}
which is precisely Eq.~\eqref{eq:POVM2} and $\overline{\Pi}_{+}+\overline{\Pi}_{-} = {\mathds{1}}$ is satisfied. The probabilities associated with these post-selected outcomes $\pm$ are
\begin{align}
    p_{\pm}={\rm Tr}[\rho \overline{\Pi}_{\pm}]=\frac{1}{2}\left[1\pm\beta\left(S_{1}\cos2\phi+S_{2}\sin2\phi\right)\right],
\end{align}
which satisfy $p_{+}+p_{-}=1$.

Compton polarimetry is efficiently described in terms of $\overline{\Pi}_{+}(\theta,\phi)$ and $\overline{\Pi}_{-}(\theta,\phi)$ as shown in Fig.~\ref{fig:setup}(b). Namely, with $N_{+}\equiv N_{\perp}$ and $N_{-}\equiv N_{\parallel}$, Eq.~\eqref{eq:ratio_compton_pol} can be written as
\begin{align}
   \dfrac{N_{+}-N_{-}}{N_{+}+N_{-}}=p_{+}-p_{-}={\rm Tr}[\rho \overline{\Pi}_{+}]-{\rm Tr}[\rho \overline{\Pi}_{-}]=\beta\left(S_{1}\cos2\phi+S_{2}\sin2\phi\right).
\end{align}


\subsection{QI Properties of The Filtered POVM Elements}
Here, we provide more details on the interpretation of $\beta$ as the parameter that determines how close the filtered POVM elements $\overline{\Pi}_{\pm}$ are to the general projector
\begin{align}
     \dyad{\psi} &= \frac{1}{2} \begin{pmatrix} \cos^2\frac{\vartheta}{2} & e^{-2i\varphi}\cos\frac{\vartheta}{2}\sin\frac{\vartheta}{2} \\ e^{2i\varphi}\cos\frac{\vartheta}{2}\sin\frac{\vartheta}{2} & \sin^2\frac{\vartheta}{2} \end{pmatrix},
\end{align}
corresponding to the general state on the Bloch sphere $\ket{\psi(\vartheta,\varphi)}=\cos(\vartheta/2)\ket{R}+e^{2i\varphi}\sin(\vartheta/2)\ket{L}$ in polarization space $(\vartheta,\varphi)$. Many of the results here can be generalized from the $N=1$ scattering case to an arbitrary number of scattering events. Thus, we omit the superscripts $(1)$ and $(N)$ where the results may be generalized. 

For convenience, in the following, we write $\overline{\Pi}_{+}(\theta,\phi)\equiv\overline{\Pi}(\theta,\phi)$ and $\overline{\Pi}_{-}(\theta,\phi)\equiv\overline{\Pi}(\theta,\phi+\pi/2)$, where

\begin{align}
    \overline{\Pi}(\theta,\phi)=\frac{1}{2}\begin{pmatrix}
        1 & \beta(\theta)e^{-2i\phi}  \\
        \beta(\theta)e^{2i\phi}  & 1
    \end{pmatrix}.
\end{align}

\subsubsection{Fidelity and Quantum State Discrimination}

The fidelity between $\overline{\Pi}$ and the general projector $\ket{\psi}\bra{\psi}$ is $F=\bra{\psi}\overline{\Pi}\ket{\psi}=\frac{1}{2}\left[1+\beta\sin(\vartheta)\cos2(\phi-\varphi)\right]$. For a single scattering event, $0\leq\beta(\theta)\leq\beta({\theta_{\text{opt}}})=0.6918$. Thus, the maximum fidelity is $F^{(1)}({\theta_{\text{opt}}})=0.8459$. We can also interpret the fidelity as the maximum success probability of the following quantum state discrimination (QSD) task.

Consider discrimination between two orthogonal states $\ket{\psi_{+}}=\cos\vartheta/2\ket{R}+e^{2i\varphi}\sin\vartheta/2\ket{L}$ and $\ket{\psi_{-}}=\sin\vartheta/2\ket{R}-e^{2i\varphi}\cos\vartheta/2\ket{L}$, which are prepared with $50\%$ probability each. Thus, the initial state of knowledge is described by the maximally mixed state $\rho=\frac{1}{2}\left(\ket{\psi_+}\bra{\psi_+}+\ket{\psi_-}\bra{\psi_-}\right)=\frac{1}{2}\mathds{1}_{2}$. The filtered POVM elements $\overline{\Pi}_{\pm}(\theta,\phi)$ describe measurement via Compton polarimetry on the initial state. Without measurement, the success probability $P_{S}$ of discriminating between states $\ket{\psi_+}$ and $\ket{\psi_-}$ is $50\%$, representing the initial state of knowledge, i.e. complete ignorance. The success probability with Compton polarimetry is given by
\begin{align}\label{eq:success}
    P_{S}&=\sum_{i=\pm}~\frac{1}{2}\Tr\left(\ket{\psi_i}\bra{\psi_i}\,\overline{\Pi}_{i}\right)\nonumber\\
    &=\frac{1}{2}\left[\bra{\psi_+}\,\overline{\Pi}_{+}\ket{\psi_+}+\bra{\psi_-}\,\overline{\Pi}_{-}\ket{\psi_-}\right]\nonumber\\
    &=\frac{1}{2}+\frac{1}{2}\left[\bra{\psi_+}\,\overline{\Pi}_{+}\ket{\psi_+}-\bra{\psi_-}\,\overline{\Pi}_{+}\ket{\psi_-}\right]\nonumber\\
    &=\frac{1}{2}\left[1+\beta\sin\vartheta\cos2(\phi-\varphi)\right],
\end{align}
is equal to $F$, and $\overline{\Pi}_{+}+\overline{\Pi}_{-}=\mathds{1}$ was used between the second and third lines. For a general pair of orthogonal states, Eq.~\eqref{eq:success} is maximized at $\phi-\varphi=m\pi$ with $m\in\mathds{Z}$, which gives $P_{S}=\frac{1}{2}\left[1+\beta(\theta_{\text{opt}})\sin\vartheta\right]$. 
This observation makes it clear that states $\ket{\psi_{+}}=\ket{R}$, $\ket{\psi_{-}}\equiv\ket{L}$ cannot be distinguished by Compton polarimetry. Furthermore, as the success probability remains $P_{S}=1/2$ at $\vartheta=0$, Compton polarimetry is blind to circular polarization.
However, orthogonal polarization states on the equator of the Bloch sphere $\vartheta=\pi/2$, corresponding to linear polarization, are maximally distinguishable. The normalized POVM elements corresponding to optimal non-projective measurements of $\{\ket{H},\ket{V}\}$ and $\{\ket{A},\ket{D}\}$ are given by 
\begin{align}
    \overline{\Pi}_{H}
  &= \frac{1}{2}\begin{pmatrix}
        1 & \beta(\theta_{\text{opt}}) \\
        \beta(\theta_{\text{opt}}) & 1
    \end{pmatrix} 
    &   \overline{\Pi}_{V}
  &= \frac{1}{2}\begin{pmatrix}
        1 & -\beta(\theta_{\text{opt}}) \\
        -\beta(\theta_{\text{opt}}) & 1
    \end{pmatrix} \nonumber \\
     \overline{\Pi}_{A}
  &= \frac{1}{2}\begin{pmatrix}
        1 & -i\beta(\theta_{\text{opt}}) \\
        i\beta(\theta_{\text{opt}}) & 1
    \end{pmatrix} 
    &   \overline{\Pi}_{D}
  &= \frac{1}{2}\begin{pmatrix}
        1 & i\beta(\theta_{\text{opt}}) \\
        -i\beta(\theta_{\text{opt}}) & 1
    \end{pmatrix},
\end{align}
which can be compared to the ideal set of projectors on linear polarization:
\begin{align}\label{eq:ideal_projectors}
    \dyad{H} &= \frac{1}{2} \begin{pmatrix} 1 & 1 \\ 1 & 1 \end{pmatrix} & 
    \dyad{V} &= \frac{1}{2} \begin{pmatrix} 1 & -1 \\ -1 & 1 \end{pmatrix} \nonumber \\[10pt]
    \dyad{A} &= \frac{1}{2} \begin{pmatrix} 1 & -i \\ i & 1 \end{pmatrix} & 
    \dyad{D} &= \frac{1}{2} \begin{pmatrix} 1 & i \\ -i & 1 \end{pmatrix},
\end{align}
which have determinant equal to zero.

\subsubsection{Trace-Norm Distance and the Helstrom Bound}
The trace-norm distance between $\overline{\Pi}$ and the general projector $\ket{\psi}\bra{\psi}$ is
\begin{align}
    D&=\frac{1}{2}\left\lVert\overline{\Pi}-\ket{\psi}\bra{\psi}\right\rVert_{1}\nonumber\\
    &=\frac{1}{2}\sum_{i}\lvert\lambda_i\rvert\nonumber\\
    &=\frac{1}{2}\sqrt{1+\beta^2-2\beta\sin\vartheta\cos2(\phi-\varphi)},
\end{align}
where $\left\lVert\ldots\right\rVert_{1}$ denotes the trace norm, $\lambda_i$ are the eigenvalues of the Hermitian matrix $\left(\overline{\Pi}-\ket{\psi}\bra{\psi}\right)$, and $D$ is minimized at $\phi-\varphi=m\pi$ and $\vartheta=\pi/2$. Thus, for a single scattering event, the minimum trace-norm distance is $D^{(1)}(\theta_{\text{opt}}=81.66^{\circ})=0.1541$.

The roles of states and measurements can be swapped to reveal a connection to the Helstrom bound as follows. Instead of utilizing the completeness relation $\overline{\Pi}_{+}+\overline{\Pi}_{-}=\mathds{1}$ between the second and third lines of Eq.~\eqref{eq:success}, if instead one utilizes $\ket{\psi_{+}}\bra{\psi_{+}}+\ket{\psi_{-}}\bra{\psi_{-}}=\mathds{1}$ the success probability reads
\begin{align}
    P^{*}_{S}&=\frac{1}{2}+\frac{1}{2}\left(\bra{\psi_+}\,\overline{\Pi}_{+}-\overline{\Pi}_{-}\ket{\psi_-}\right).
\end{align}
We denote this success probability $P^{*}_{S}$ to emphasize its distinct operational interpretation compared to $P_{S}$. Namely, the roles of POVM elements and states have been swapped. Here, $\overline{\Pi}_{+}$ and $\overline{\Pi}_{-}$ are two states to be distinguished, and $\ket{\psi_{+}}\bra{\psi_{+}}$ and $\ket{\psi_{-}}\bra{\psi_{-}}$ are a complete set of projective POVMs to optimize over. Note that this is not the measurement scenario described by Compton polarimetry, but we consider this scenario to provide further insight into $\beta$. Optimizing over $\vartheta$ and $\varphi$ gives 
\begin{align}
    P^{*}_{S}&=\frac{1}{2}\left\{1+\max_{\vartheta,\varphi}\,\Tr\left[\ket{\psi_{+}}\bra{\psi_{+}}\left(\overline{\Pi}_{+}-\overline{\Pi}_{-}\right)\right]\right\}\nonumber\\
    &=\frac{1}{2}\left[1+D\left(\overline{\Pi}_{+},\overline{\Pi}_{-}\right)\right]\nonumber\\
    &=\frac{1}{2}\left[1+\beta(\theta)\right],
\end{align}
which is precisely the Helstrom bound for this measurement scenario. Notably, we have used $D\left(\overline{\Pi}_{+},\overline{\Pi}_{-}\right)=\beta(\theta)$. In this way, we see that $\beta$ also represents the maximum distinguishability between $\overline{\Pi}_{+}$ and $\overline{\Pi}_{-}$.


\subsection{Generalization to Sequential Scattering Events}

\subsubsection{$N$-fold Cross Section}

The $N$-fold differential scattering cross section for the co-planar trajectories considered in the main text is
\begin{align}\label{eq:N-fold-DSC_2}
    \left. \dfrac{d^N\sigma}{d\Omega_1 d\Omega_2\ldots d\Omega_N}\right\rvert_{\text{co}}&=\bra{I}T_{N}\,T_{N-1}\ldots T_{3}\,T_{2}\,T_{1}M(\phi)\,\ket{S},
\end{align}
where $\ket{S}$ is the Stokes vector for an arbitrary initial polarization state $\rho$ with $S_0=1$. Note that $T_{N}\,T_{N-1}\ldots T_{3}\,T_{2}\,T_{1}M(\phi)\,\ket{S}$ is the Stokes vector that represents the unnormalized state after sequential scatterings through the co-planar trajectories specifically, and that Eq.~\eqref{eq:N-fold-DSC} applies to arbitrary sequential scattering events. As each $T_{j}$ matrix is block diagonal, the matrix product $T_{N}T_{N-1}\ldots T_{2}T_{1}$ will also be block diagonal and of the form
\begin{align}
   T_{N}T_{N-1}\ldots T_{2}T_{1}&= r_e^{2N} \begin{pmatrix}
        \alpha(\bm{\theta}) & \gamma(\bm{\theta}) & 0 & 0\\
        \delta(\bm{\theta}) & \epsilon(\bm{\theta}) & 0 & 0 \\
        0 & 0 & f(\bm{\theta}) & 0 \\
        0 & 0 & 0 & g(\bm{\theta})
    \end{pmatrix},
\end{align}
where the elements of this matrix are functions of the $N$ polar angles $\bm{\theta}=(\theta_1,\theta_2,\ldots,\theta_N)$. Multiplying this block diagonal matrix from the right by the rotation matrix $M(\phi)$, gives 
\begin{align}\label{eq:trans_matrix_particular_trajectory}
   T_{N}T_{N-1}\ldots T_{2}T_{1}M(\phi)&= r_e^{2N} \begin{pmatrix}
        \alpha(\bm{\theta}) &  \gamma(\bm{\theta})\cos2\phi & \ \gamma(\bm{\theta})\sin2\phi & 0\\
        \delta(\bm{\theta}) & \epsilon(\bm{\theta})\cos2\phi & \epsilon(\bm{\theta})\sin2\phi & 0 \\
        0 & -f(\bm{\theta})\sin2\phi & f(\bm{\theta})\cos2\phi & 0 \\
        0 & 0 & 0 & g(\bm{\theta})
    \end{pmatrix}.
\end{align}
Inserting this matrix product into Eq.~\eqref{eq:N-fold-DSC_2} evaluates to
\begin{align}\label{eq:N-fold-DSC_3}
    \left. \dfrac{d^N\sigma}{d\Omega_1 d\Omega_2\ldots d\Omega_N}\right\rvert_{\text{co}}&=r_e^{2N}\alpha(\bm{\theta})\left\{1+\beta(\bm{\theta})\left( S_1 \cos 2\phi + S_2 \sin 2\phi \right)\right\}
\end{align}
where $\gamma(\bm{\theta})= \alpha(\bm{\theta})\beta(\bm{\theta})$. Therefore, only the functions $\alpha$ and $\gamma= \alpha\beta$ appear in the final cross section and we compute these numerically via the Compton transition matrix elements listed in the End Matter.

\subsubsection{POVM}

For an arbitrary trajectory of sequential scatterings, the Stokes vector is $T_{N}M_{N}\ldots T_{2}M_{2}\,T_{1}M_{1}\,\ket{S}$ and the $N$-fold differential scattering cross section is given by Eq.~\eqref{eq:N-fold-DSC} in the main text.  The total cross section $\sigma^{(N)}_{\text{tot}}$ is calculated by integrating Eq.~\eqref{eq:N-fold-DSC} over all solid angles
\begin{align}
\sigma^{(N)}_{\text{tot}}&=\int ~ d\Omega_{1}~\int ~ d\Omega_{2}\ldots\int ~ d\Omega_{N}~\dfrac{d^N\sigma}{d\Omega_1 d\Omega_2\ldots d\Omega_N},
\end{align}
and the cross sections up to $N=5$ are given in Table~\ref{tab:sigma_tot}. We calculate these values numerically and note that one may use
\begin{align}
\int_{0}^{2\pi}\,d\phi_{j} \, M(\phi_{j})=2\pi\begin{pmatrix}
    1 & 0 & 0 & 0 \\
    0 & 0 & 0 & 0 \\
    0 & 0 & 0 & 0 \\
    0 & 0 & 0 & 1
\end{pmatrix},
\end{align}
to simplify the numerical integration.

\begin{table}[h]
\centering
\caption{Total $N$-fold cross sections up to $N = 5$ for $E_0 = 1.0$. Note the total phase space grows with each successive scattering.}
\label{tab:sigma_tot}
\begin{tabular}{c c}
\toprule
$N$ & Total Cross Section $\sigma_{tot}^{(N)} / r_e^{2N}$ \\ \midrule
1 & 3.60846 \\ 
2 & 15.76223 \\ 
3 & 78.91078 \\ 
4 & 434.75406 \\ 
5 & 2564.81279 \\ 
\bottomrule\end{tabular}
\end{table}

As for the $N=1$ case, the probability density for scattering along an arbitrary trajectory ${d^{N} p (\bm{\theta},\bm{\phi})}$ is found by normalizing the differential cross section by $\sigma^{(N)}_{\text{tot}}$. The POVM elements are then defined implicitly via
\begin{align}\label{eq:probN}
    \frac{d^{N} p (\bm{\theta},\bm{\phi})}{{d\Omega_1 d\Omega_2\ldots d\Omega_N}}&=\Tr\left(\rho ~\Pi^{(N)}(\bm{\theta},\bm{\phi})\right).
\end{align}
Focusing on the co-planar trajectories, the POVM elements ${\Pi}^{(N)}(\bm{\theta},\phi)$ may be deduced in Section~\ref{sec:supp_deducing}, namely by using Eqs~\eqref{eq:N-fold-DSC_3} and \eqref{eq:probN} and thus equating 
\begin{align}
r_e^{2N}\dfrac{\alpha(\bm{\theta})}{\sigma^{(N)}_{\text{tot}}}\left\{1+\beta(\bm{\theta})\left( S_1 \cos 2\phi + S_2 \sin 2\phi \right)\right\}=\Tr\left(\rho ~\Pi^{(N)}(\bm{\theta},{\phi})\right)=\frac{1}{2}\sum_{j}S_j {\rm Tr}\left(\sigma_j \Pi^{(N)}(\bm{\theta},\phi)\right).
\end{align}
This procedure gives
 \begin{align}
    {\Pi}^{(N)}(\bm{\theta},\phi)=r_e^{2N}\,\dfrac{\alpha(\bm{\theta})}{\sigma^{(N)}_{\text{tot}}}\begin{pmatrix}
        1 & \beta(\bm{\theta})e^{-2i\phi} \\
        \beta(\bm{\theta})e^{2i\phi} & 1
    \end{pmatrix}
\end{align}
and the dichotomic, normalized polarization POVM elements in Eq.~\eqref{eq:POVMN} follow from the same analysis as in Section~\ref{sec:supp_filtering}.

\subsubsection{The QI properties of the Filtered POVM Elements}
As the POVM elements given by Eq.~\eqref{eq:POVM2} and Eq.~\eqref{eq:POVMN} are of the same form, the QI properties of the normalized POVM elements $\overline{\Pi}(\bm{\theta},{\phi})$ generalize from those of $\overline{\Pi}({\theta},{\phi})$. For example, 
\begin{align}
 &F^{(N)}=P_{S}=\bra{\psi}\overline{\Pi}(\bm{\theta},{\phi})\ket{\psi}=\frac{1}{2}\left[1+\beta(\bm{\theta})\sin(\vartheta)\cos2(\phi-\varphi)\right],   \\
 &D^{(N)}=\frac{1}{2}\left\lVert\overline{\Pi}(\bm{\theta},{\phi})-\ket{\psi}\bra{\psi}\right\rVert_{1}=\frac{1}{2}\sqrt{1+\beta^2(\bm{\theta})-2\beta(\bm{\theta})\sin\vartheta\cos2(\phi-\varphi)}, \\
 &D\left(\overline{\Pi}_{+}(\bm{\theta},{\phi}),\overline{\Pi}_{-}(\bm{\theta},{\phi})\right)=\beta(\bm{\theta}),
\end{align}
all continue to hold. Also, the POVM elements corresponding to measurements in the horizontal/vertical and diagonal/anti-diagonal bases are
\begin{align}
    \overline{\Pi}_{H}
  &= \frac{1}{2}\begin{pmatrix}
        1 & \beta(\bm{\theta}) \\
        \beta(\bm{\theta}) & 1
    \end{pmatrix} 
    &   \overline{\Pi}_{V}
  &= \frac{1}{2}\begin{pmatrix}
        1 & -\beta(\bm{\theta}) \\
        -\beta(\bm{\theta}) & 1
    \end{pmatrix} \nonumber \\
     \overline{\Pi}_{A}
  &= \frac{1}{2}\begin{pmatrix}
        1 & -i\beta(\bm{\theta}) \\
        i\beta(\bm{\theta}) & 1
    \end{pmatrix} 
    &   \overline{\Pi}_{D}
  &= \frac{1}{2}\begin{pmatrix}
        1 & i\beta(\bm{\theta}) \\
        -i\beta(\bm{\theta}) & 1
    \end{pmatrix}.
\end{align}
By optimizing over $\bm{\theta}$, and in the limit of a high number of scattering events, we observe $\beta\rightarrow 1$ and these POVM elements tend to ideal projective measurements of linear polarization---see Eq.~\eqref{eq:ideal_projectors}.


\section{Further Details on the Bell Test}

\subsection{Derivation of the CHSH function}\label{supp:derivation_CHSH}
Here, we provide further details on the derivation of Eq.~\eqref{eq:Sfunc_N} in the main text. Firstly, in Section~\ref{sec:short_derivation}, we present an efficient derivation using the filtered POVM elements. Following this, in Section~\ref{sec:long_derivation}, we present a derivation of the CHSH function in terms of the POVM elements $\Pi^{(N)}(\bm{\theta},\phi)$. This presentation enables an exploration of the CHSH function for a range of angular deviations about optimal scattering angles, which is relevant for the following section on simulations.

\subsubsection{Efficient Calculation via Filtered POVM Elements}\label{sec:short_derivation}

Denoting the post-selected joint probabilities on the right-hand side of Eq.~\eqref{eq:Prob_phia_phib} as
\begin{align}
p_{jk}(\phi_a,\phi_b)&=\bra{\Phi^{-}}\overline{\Pi}_{j}(\phi_a)\otimes\overline{\Pi}_{k}(\phi_b)\ket{\Phi^{-}}\nonumber\\
&=\frac{1}{4}\left[1-(jk)\,\beta^2(\bm{\theta})\,\cos2(\phi_a+\phi_b)\right],
\end{align}
where $j,k=\pm$, then Eq.~\eqref{eq:eval_exp} evaluates to
\begin{align}\label{eq:supp_exp_val}
    E(\phi_a,\phi_b)&=p_{++}(\phi_a,\phi_b)+p_{--}(\phi_a,\phi_b)-p_{+-}(\phi_a,\phi_b)-p_{-+}(\phi_a,\phi_b)\nonumber\\
    &=-\beta^2(\bm{\theta})\,\cos2(\phi_a+\phi_b).
\end{align}
Here, $E(\phi_a,\phi_b)$ has exactly the form of the standard quantum mechanical prediction for a maximally entangled state measured with ideal projectors, but with $\beta^2$ playing the role of the visibility. 
Note that Eq.~\eqref{eq:supp_exp_val} is valid for any value of $\phi_a$ and $\phi_b$ Alice and Bob may choose, including their second measurement settings $\phi_{a'}$ and $\phi_{b'}$.
The CHSH function $S$ depends on 4 azimuthal angles corresponding to Alice's $\{\phi_a,\phi_{a'}\}$ and Bob's $\{\phi_b,\phi_{b'}\}$ measurement settings. However, if we choose these settings to correspond to the Bell-test angles, the expression for the CHSH function in the main text follows directly from Eq.~\eqref{eq:supp_exp_val}:
\begin{align}
    S(\bm{\theta},\phi)&=E(0,-\phi)-E(0,-3\phi)+E(2\phi,-\phi)+E(2\phi,-3\phi)\nonumber\\
    &=\beta^2(\bm{\theta})\left(-3\cos2\phi+\cos6\phi\right).
\end{align}

\subsubsection{Derivation via Unfiltered POVM Elements}\label{sec:long_derivation}

Using the relation $\Pi^{(N)}(\bm{\theta},\phi)=e^{-i\phi \sigma_{z}}\Pi^{(N)}(\bm{\theta},0)e^{i\phi \sigma_{z}}$, the joint scattering probability density may be written as
\begin{align}\label{eq:supp_prob_1}
    \text{Prob}(\phi_{a},\phi_{b})&=\Tr\left[\ket{\Phi^{-}}\bra{\Phi^{-}}\,e^{-i\phi_a \sigma_{z}}\Pi^{(N)}(\bm{\theta},0)e^{i\phi_a \sigma_{z}}\otimes e^{-i\phi_b \sigma_{z}}\Pi^{(N)}(\bm{\theta},0)e^{i\phi_b \sigma_{z}}\right]\\
    &=\Tr\left[\rho(\phi_a,\phi_b)\,\Pi^{(N)}(\bm{\theta},0)\otimes\Pi^{(N)}(\bm{\theta},0)\right],
\end{align}
where we have used the cyclic property of the trace to absorb the local unitaries into a rotated state $\rho(\phi_a,\phi_b)=U^{\dagger}(\phi_a,\phi_b)\ket{\Phi^{-}}\bra{\Phi^{-}}U(\phi_a,\phi_b)$ and $U(\phi_a,\phi_b)=e^{-i\phi_a \sigma_{z}}\otimes e^{-i\phi_b \sigma_{z}}$. In the circular basis, 
\begin{align}
    \rho(\phi_a,\phi_b)=\rho(\phi_a+\pi/2,\phi_b+\pi/2)&=\frac{1}{2}\begin{pmatrix}
        1 & 0 & 0 & -e^{i(2\phi_a+2\phi_b)} \\
        0 & 0 & 0 & 0 \\
        0 & 0 & 0 & 0 \\
        -e^{-i(2\phi_a+2\phi_b)} & 0 & 0 & 1
    \end{pmatrix},\\
    \rho(\phi_a,\phi_b+\pi/2)=\rho(\phi_a+\pi/2,\phi_b)&=\frac{1}{2}\begin{pmatrix}
        1 & 0 & 0 & e^{i(2\phi_a+2\phi_b)} \\
        0 & 0 & 0 & 0 \\
        0 & 0 & 0 & 0 \\
        e^{-i(2\phi_a+2\phi_b)} & 0 & 0 & 1
    \end{pmatrix},
\end{align}
and
\begin{align}\label{eq:supp_tensor_prod}
    \Pi^{(N)}(\bm{\theta},0)\,\otimes\,\Pi^{(N)}(\bm{\theta},0)
    &=\mathcal{N}^2\begin{pmatrix}
        1 & \beta & \beta & \beta^2 \\
        \beta & 1 & \beta^2 & \beta \\
        \beta & \beta^2 & 1 & \beta \\
        \beta^2 & \beta & \beta & 1
        \end{pmatrix},
\end{align}
where we have suppressed the $\bm{\theta}$ argument on the right-hand side, and we use $\mathcal{N}=r_e^{2N}\,{\alpha}/{\sigma_{\text{tot}}^{(N)}}$---generalizing this quantity from the $N=1$ case, for tidiness. Note that $\rho(\phi_a,\phi_b)=\rho(\phi_a+\pi/2,\phi_b+\pi/2)$ reflects the symmetry of the state $\ket{\Phi^{-}}$ under simultaneous local $\pi/2$ rotations.

For a large number of measurements, the coincidence count rates in Eq.~\eqref{eq:eval_exp} may be replaced by their associated probabilities. Explicitly, 
\begin{align}\label{eq:supp_eval_exp}
    E(\phi_a,\phi_b)=\dfrac{\text{Prob}(\phi_{a},\phi_{b})+\text{Prob}(\phi_{a}+\pi/2,\phi_{b}+\pi/2)-\text{Prob}(\phi_{a}+\pi/2,\phi_{b})-\text{Prob}(\phi_{a},\phi_{b}+\pi/2)}{\text{Prob}(\phi_{a},\phi_{b})+\text{Prob}(\phi_{a}+\pi/2,\phi_{b}+\pi/2)+\text{Prob}(\phi_{a}+\pi/2,\phi_{b})+\text{Prob}(\phi_{a},\phi_{b}+\pi/2)}.
\end{align}
Using Eqs.~\eqref{eq:supp_prob_1} to \eqref{eq:supp_tensor_prod}, the probability densities in $ E(\phi_a,\phi_b)$ are
\begin{align}
    \text{Prob}(\phi_a,\phi_b) = \text{Prob}(\phi_a+\pi/2,\phi_b+\pi/2) & =\mathcal{N}^2\left[1-\beta^2\,\cos2(\phi_a+\phi_b)\right],\label{eq:supp_unfiltered_prob1}\\
    \text{Prob}(\phi_a,\phi_b+\pi/2) = \text{Prob}(\phi_a+\pi/2,\phi_b) &= \mathcal{N}^2\left[1+\beta^2\,\cos2(\phi_a+\phi_b)\right].\label{eq:supp_unfiltered_prob2}
\end{align}
and hence
\begin{align}\label{eq:supp_exp_val_2}
    E(\phi_a,\phi_b)=-\beta^2(\bm{\theta})\,\cos2(\phi_a+\phi_b).
\end{align}
By selecting the Bell-test angles, the expression for the CHSH function in the main text immediately follows.

\subsection{Simulations of Total Coincidence Count Rate}

The kinematics of the $N=2$ case is simulated using the \textsc{Geant4} package. The simulation starts by generating two $511\,\text{keV}$ photons of orthogonal linear polarization travelling in opposite directions that lie within $10^{\circ}$ from the $z$ axis. Since \textsc{Geant4} propagates each photon separately, the nature of these photons is fully classical, i.e. no entangled correlations. It is worth noting that the built-in quantum entanglement option for photons in \textsc{Geant4} simply enforces a Pryce--Ward cross section when there are two Compton scatterings~\cite{watts2021photon,zugec2026reconciliation}, and is, thus, not applicable to the $N = 2$ case under consideration. As a result, this simulation is essentially a four-fold Klein--Nishina description of this quadruple Compton scattering scenario.

The geometrical construction of this simulation consists of two pairs of Compton scatterers made of LYSO crystals. The first pair of scatterers, placed right next to the source of emission centred at the origin, are two cuboids that are $2\,\text{cm}$ thick along the $z$ axis with a $1\,\text{cm}$ square cross section. Roughly $5\,\text{cm}$ away are the second pair of scatterers which take the shape of annular rings of $15\,\text{cm}$ diameter and $2\,\text{cm}$ thickness centred along the $z$ axis. 
Note that owing to the azimuthal symmetry of the initial two-photon state, data can be accumulated over all azimuthal angles of Alice and Bob's detectors. In other words, the choice of $\phi_a=0$ in the main text is arbitrary. With rotated Bell-test-angle settings, $\phi_a\neq0$, $\phi_{a'}=2\phi+\phi_a$, $\phi_b=-\phi-\phi_a$, and $\phi_{b'}=-3\phi-\phi_a$, the CHSH function is still given by $S(\bm{\theta},\phi)=\beta^2(\bm{\theta})\left(-3\cos2\phi+\cos6\phi\right)$
Further note that these quoted dimensions of the scatterers are, to some extent, arbitrary as long as they cover the angular region where a CHSH violation may still be observed. To establish this region, the CHSH function may be explored away from its optimal scattering angles numerically via Eqs~\eqref{eq:supp_unfiltered_prob1} and \eqref{eq:supp_unfiltered_prob2}.

In the desired scenario, each simulated photon will undergo exactly one Compton scatter in each scatterer at angles listed in Table~\ref{tab:optimal_angles}, within an angular tolerance still capable of violating a CHSH inequality. The count rate of a single photon, per simulated photon pair, that satisfy this double-scattering criterion was then found to be $C_1 = 5.6\times10^{-6}$, and, hence, the case for such scatterings on both sides becomes $C_2 = C_1^2 = 3.1\times 10^{-11}$. Taking the initial angular acceptance into account, this would result in a per positron-electron annihilation, or per source decay, count rate of $\mathcal{C}/r=\text{5}\times10^{-13}$.

\end{document}